\documentclass[sn-mathphys]{sn-jnl}

\usepackage{graphicx}
\usepackage{xcolor}
\usepackage{pdfpages}
\usepackage{url}
\usepackage{amssymb,stackengine,graphicx}

\jyear{2021}%
\begin{document}

\title{Hybrid Equations of State for Neutron Stars with Hyperons and Deltas}

\author{\fnm{A.} \sur{Clevinger}}
\author{\fnm{J.} \sur{Corkish}}
\author{\fnm{K.} \sur{Aryal}}
\author{\fnm{V.} \sur{Dexheimer
}}

\affil{\orgdiv{Department of Physics},  \orgname{Kent State University},  \orgaddress{\city{Kent} \state{OH}, \postcode{44242}, \country{USA}}}

\abstract
{In this contribution, we describe new chemically-equilibrated charge-neutral hybrid equations of state for neutron stars. They present a first-order phase transition to quark matter and differentiate by the particle population considered and how these particles interact. While some equations of state contain just nucleons and up, down-quarks, others also contain hyperons, Delta baryons, and strange quarks. The hybrid equations of state, together with corresponding hadronic ones, are available on the CompOSE repository and can be used for different astrophysical applications.
}

\maketitle


\section{Introduction}

The extreme conditions that give birth to neutron stars, i.e. the gravitational collapse of massive stars that can no longer perform fusion, are the perfect laboratory to create new phases of matter. Such extreme conditions of density (with comparatively low temperature) cannot be replicated in laboratories on Earth and are, therefore, very difficult to constrain. Mathematically, the core of neutron stars, which occupies $\sim 90\%$ of their radii \cite{2008LRR....11...10C}, can be modelled by assuming infinite bulk baryonic matter strongly interacting through a mean field (in space and time) of mesons \cite{WALECKA1974491}. These baryons go beyond ordinary nucleons (protons and neutrons), and include hyperons (other members of the spin $1/2$ octet that contain strange quarks) and the four Delta's ($\Delta$'s, the lighter resonances of the spin $3/2$ decuplet), all of which are more massive than the nucleons and, as a consequence, appear at larger densities. The remaining resonances are too heavy and do not appear inside neutron stars within our formalism. In addition, at several times nuclear saturation density, baryons start to overlap \cite{BAYM1979131}, a description of matter based exclusively on baryons loses its meaning, and the possibility of a phase transition to quark matter must be considered.

In this work, we discuss 16 equations of state (EoS's) that are produced assuming different particle composition (8 hadronic and 8 hybrid) and different vector isoscalar and vector isovector (repulsive) interaction terms. We make use of a chiral model in which the scalar (attractive) terms are fixed in order to reproduce the vacuum masses of hadrons (baryons and mesons), as well as mesonic decay constants. While vector terms are constrained to reproduce isospin-symmetric matter saturation properties measured in the laboratory, there is still room to vary isovector terms and higher-order vector terms, which do not affect significantly saturation properties, but can be important for astrophysics \cite{Pais:2016xiu,Ribes:2019kno,Christian:2019qer,Marczenko:2020jma,Lopes:2020btp,Ferreira:2021yje}.

Recently, a $3.1\,\sigma$ measurement of a critical point was performed at RHIC in the Beam Energy Scan experiment \cite{STAR:2020tga}. For the moment, more data is being collected to increase significance, but this would imply that a first-order phase transition exists for a given temperature, in which case (unless proven otherwise) we can assume that it extends all the way to zero temperature. We analyze in this work in detail the effect of a first-order phase transition to quark matter in different EoS's at zero temperature. We discuss the resulting particle populations within the relevant density regime and the corresponding mass and radius of the neutron stars they generate. Finally, we discuss the common-format EoS tables that are available for the astrophysics community on the CompOSE repository \cite{Oertel:2016bki,Typel:2013rza,compose}. 

\section{Formalism}

To describe strongly interacting bulk matter in the core of neutron stars, we make use of the Chiral Mean Field (CMF) model \cite{Papazoglou:1998vr,Dexheimer:2008ax}, which is an effective relativistic model that includes the two important features predicted by QCD to take place at large density, namely chiral symmetry restoration and deconfinement to quark matter. As the density increases, nucleons ($p$, $n$) give space to the rest of the baryon octet ($\Lambda$, $\Sigma^+$, $\Sigma^0$, $\Sigma^-$, $\Xi^0$, $\Xi^-$), baryons from the spin $3/2$ decuplet ($\Delta^{++}$, $\Delta^+$, $\Delta^0$, $\Delta^-$, $\Sigma^{*+}$, $\Sigma^{*0}$, $\Sigma^{*-}$, $\Xi^{*0}$, $\Xi^{*-}$, $\Omega$), and, eventually, to the three light quarks ($u$, $d$, $s$). All of these are described by the same model, being baryons favoured at low densities and quarks at high densities due to the way their effective masses are derived, including a dependence on a field $\Phi$ associated with the Polyakov loop \cite{Dexheimer:2009hi}. In our description, deconfinement to quark matter takes place at zero temperature in the form of a first-order phase transition, so there is a jump in (baryon number) density associated with a jump in $\Phi$. Please refer to Refs.~\cite{Roark:2018boj} for relevant equations and a complete list of coupling constants.

In this work, we use a modified version of the quark description from Refs.~\cite{Dexheimer:2009hi,Roark:2018boj}, recently introduced in Ref.~\cite{Dexheimer:2020rlp}, which reproduces weaker first-order phase transitions (with smaller density jumps) and allows for stable hybrid stars without having to resort to mixtures of phases.
The modifications include a change in the field $\Phi$ potential to depend less (lower power) on the baryon chemical potential, a different strength of coupling between baryons/quarks and $\Phi$, a lower value of scalar quark couplings, and new vector quark couplings.
Here, for the first time, we apply this modified quark description to several different particle compositions and interactions. These additional (strong force) interactions were shown to improve the agreement of hadronic models with laboratory and observational constraints \cite{Horowitz:2002mb,Dexheimer:2018dhb,Dexheimer:2020rlp}. The resulting EoS's are hereforth referred to by number:

\vspace{.2cm}
\begin{tabular}{ll}
\textbf{\textasteriskcentered\ EoS 1}&with standard interactions\\
&\textbf{hadronic:} nucleons, hyperons, electrons, and muons\\ 
&\textbf{hybrid:} nucleons, hyperons, uds quarks, electrons, and muons\\ &with phase transition at $n_{B}$ = 0.472 fm$^{-3}$
\vspace{.2cm}\\
\textbf{\textasteriskcentered\ EoS 2}&with standard interactions\\
&\textbf{hadronic:} nucleons and electrons\\
&\textbf{hybrid:} nucleons, ud quarks, and electrons\\ &with phase transition at $n_{B}$ = 0.433 fm$^{-3}$
\vspace{.2cm}\\
\textbf{\textasteriskcentered\ EoS 3}&with $\omega\rho$ terms\\
&\textbf{hadronic:} nucleons, hyperons, electrons, and muons\\
&\textbf{hybrid:} nucleons, hyperons, uds quarks, electrons, and muons\\ &with phase transition at $n_{B}$ = 0.638 fm$^{-3}$
\vspace{.2cm}\\
\textbf{\textasteriskcentered\ EoS 4}&with $\omega\rho$ terms\\
&\textbf{hadronic:} nucleons and electrons\\
&\textbf{hybrid:} nucleons, ud quarks, and electrons\\ &with phase transition at $n_{B}$ = 0.561 fm$^{-3}$
\vspace{.2cm}\\
\textbf{\textasteriskcentered\ EoS 5}&with $\omega\rho$ and $\omega^4$ terms\\
&\textbf{hadronic:} nucleons, hyperons, electrons, and muons\\
&\textbf{hybrid:} nucleons, hyperons, uds quarks, electrons, and muons\\ &with phase transition at $n_{B}$ = 0.688 fm$^{-3}$
\vspace{.2cm}\\
\textbf{\textasteriskcentered\ EoS 6}&with $\omega\rho$ and $\omega^4$ terms\\
&\textbf{hadronic:} nucleons and electrons\\ 
&\textbf{hybrid:} nucleons, ud quarks, and electrons\\ &with phase transition at $n_{B}$ = 0.629 fm$^{-3}$
\vspace{.2cm}\\
\textbf{\textasteriskcentered\ EoS 7}&with $\omega\rho$ and $\omega^4$ terms\\
&\textbf{hadronic:} nucleons, hyperons, $\Delta$'s, electrons, and muons\\
&\textbf{hybrid:} nucleons, hyperons, $\Delta$'s, uds quarks, electrons, and muons\\
&with phase transition at $n_{B}$ = 0.689 fm$^{-3}$
\end{tabular}

\begin{tabular}{ll}
\textbf{\textasteriskcentered\ EoS 8}&with $\omega\rho$ and $\omega^4$ terms\\
&\textbf{hadronic:} nucleons, $\Delta$'s, and electrons\\ 
&\textbf{hybrid:} nucleons, $\Delta$'s, ud quarks, and electrons\\ &with phase transition at $n_{B}$ = 0.644 fm$^{-3}$
\end{tabular}
\\

where the term standard refers to the interactions previously used in all CMF model publications \cite{Roark:2018uls}. The introduction of the separate $\omega^4$ coupling implies that nucleonic $\omega$-meson couplings have to be refit to reproduce saturation properties. Our $\omega^4$ coupling strength value was chosen to be as large as possible, with the limitation that the compressibility for isospin-symmetric matter at saturation does not increase. In this case, the quark and $\Phi$ couplings are also modified. See Ref.~\cite{Dexheimer:2020rlp} for details. The $\omega\rho$ coupling does not affect saturation, and our coupling strength value was chosen to be as large as possible, with the limitation that the mass of hadronic stars still reaches $2$ M$_\odot$ in the absence of the $\omega^4$ coupling. Note that not all the particle species specified above appear for zero temperature in the regime relevant for neutron stars. From the baryon octet, the $\Xi$ never appears, and from the decuplet, the $\Sigma^*$, $\Xi^*$, and $\Omega$ never appear. Our result is orthogonal with the ones from Refs.~\cite{RIKOVSKASTONE2007341,Weissenborn:2011ut}, where neutron stars with $\Xi$'s, but no $\Sigma$'s are predicted. Such differences between models are connected to their optical potentials for the hyperons, which even at saturation are still widely unknown (except for the $\Lambda$ \cite{PhysRevC.38.2700}). In particular, in Refs.~~\cite{RIKOVSKASTONE2007341,Weissenborn:2011ut}, the $\Sigma$ potential (for isospin symmetric matter at saturation) is very repulsive, opening space for the appearance of $\Xi$'s.
In the CMF model, there is only one coupling related to explicit symmetry breaking that can be fitted to the $\Lambda$ potential. Work on going beyond SU(3) and SU(6) symmetries for the couplings is underway and is expected to produce changes in the hyperon potentials. 

\section{Results}

We discuss first the particle population generated by the $16$ EoS's (8 hadronic and 8 hybrid). They describe cold ($T=0$) chemically-equilibrated charge-neutral neutron-star matter, which contains baryons, quarks (the hybrid ones), electrons, and muons (in the EoS's with hyperons). Hadronic and quark charge fraction or, equivalently, leptonic fraction is determined self-consistently by the imposition of charge neutrality, together with chemical equilibrium. The standard and additional vector interactions discussed in the Formalism section are present in both hadronic and quark phases (when included).

The nuclear properties we reproduce for isospin-symmetric matter are: saturation density $n_B=0.15$ fm$^{-3}$, binding energy per nucleon $B=-16$ MeV, compressibility $K=300$ MeV, and symmetry energy $E_{\rm{sym}}=30$ MeV. Hyperon potentials for symmetric matter at saturation are $U_\Lambda=-28$ MeV, $U_\Sigma=5$ MeV, $U_\Xi=-18$ MeV ($U_\Lambda=-27$ MeV, $U_\Sigma=6$ MeV, $U_\Xi=-17$ MeV, $U_\Delta=-64$ MeV when $\omega^4$ interactions are included). The symmetry energy slope is $L=88$ MeV ($L=75$ MeV when $\omega\rho$ interactions are included).

\begin{figure*}[t!]
  \includegraphics[width=1.01\linewidth]{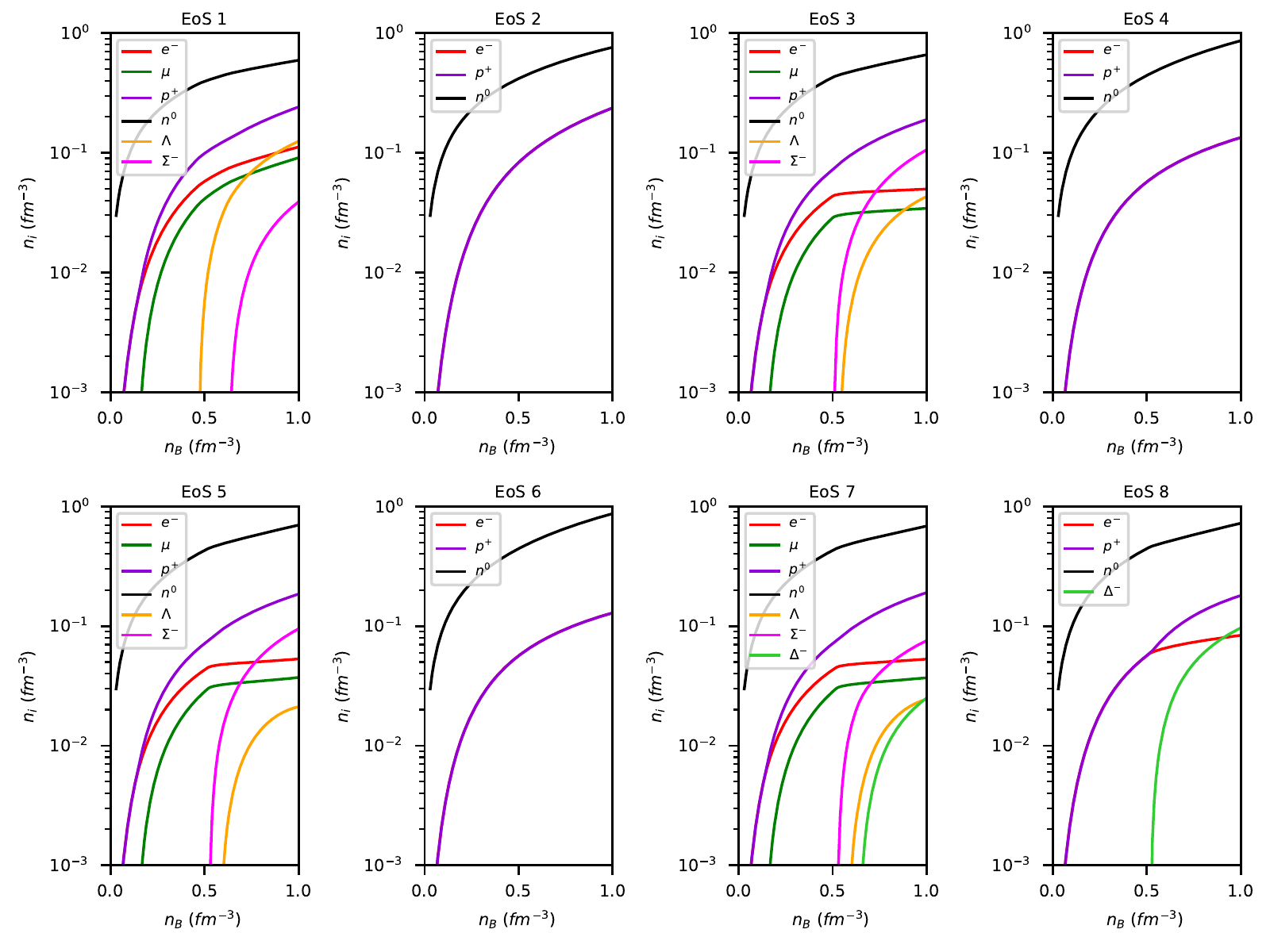}
  \caption{Particle population as a function of density for different hadronic equations of state. For EoS's 2, 4, and 6, the line for the electrons coincides with the one for the protons.}
  \label{fig:1}
\end{figure*}

It can be seen for EoS 1 in Fig.~\ref{fig:1} how the only hyperons present in our calculations are the lighter $\Lambda$ and the negatively-charged $\Sigma^-$ (enhanced due to the charge neutrality constraint). In EoS 2, hyperons and muons are not included. In EoS 3, the additional $\omega\rho$ interaction decreases the cost of producing isospin asymmetry, having the direct effect of decreasing the asymmetry energy slope $L$ \cite{Horowitz:2001ya,Logoteta:2013ipa,Hu:2020ujf}, thus increasing the asymmetry between protons and neutrons  and modifying the hyperon amount and order of appearance. In EoS 4, hyperons and muons are not included, but the isospin difference can once more be seen. For EoS 5, the additional $\omega^4$ interaction changes the particle population only at large density. For EoS 7, the only baryon with spin $3/2$ that appears in the $\Delta^-$ (enhanced due to charge neutrality) in comparable amount with the $\Lambda$'s. In EoS 8, without the hyperons, the $\Delta$'s appear in significant amount. Notice that we keep fixed the largely unknown vector coupling of the $\Delta$ baryon \cite{Kolomeitsev:2016ptu,Cozma:2021tfu}. We fix the $\Delta$ vector coupling to one value ($g_{\Delta\omega}/g_{N\omega}=1.25$) but, if we were to vary it, this would increase or decrease the $\Delta$ amount, to the extent that they would suppress all hyperons (EoS 8) or disappear (EoS 5). In this sense, we covered all the possibilities in this figure. The scalar couplings are, as discussed before, constrained in chiral models to reproduce vacuum masses, which also applies to the $\Delta$'s.

\begin{figure*}[t!]
  \includegraphics[width=1.01\linewidth]{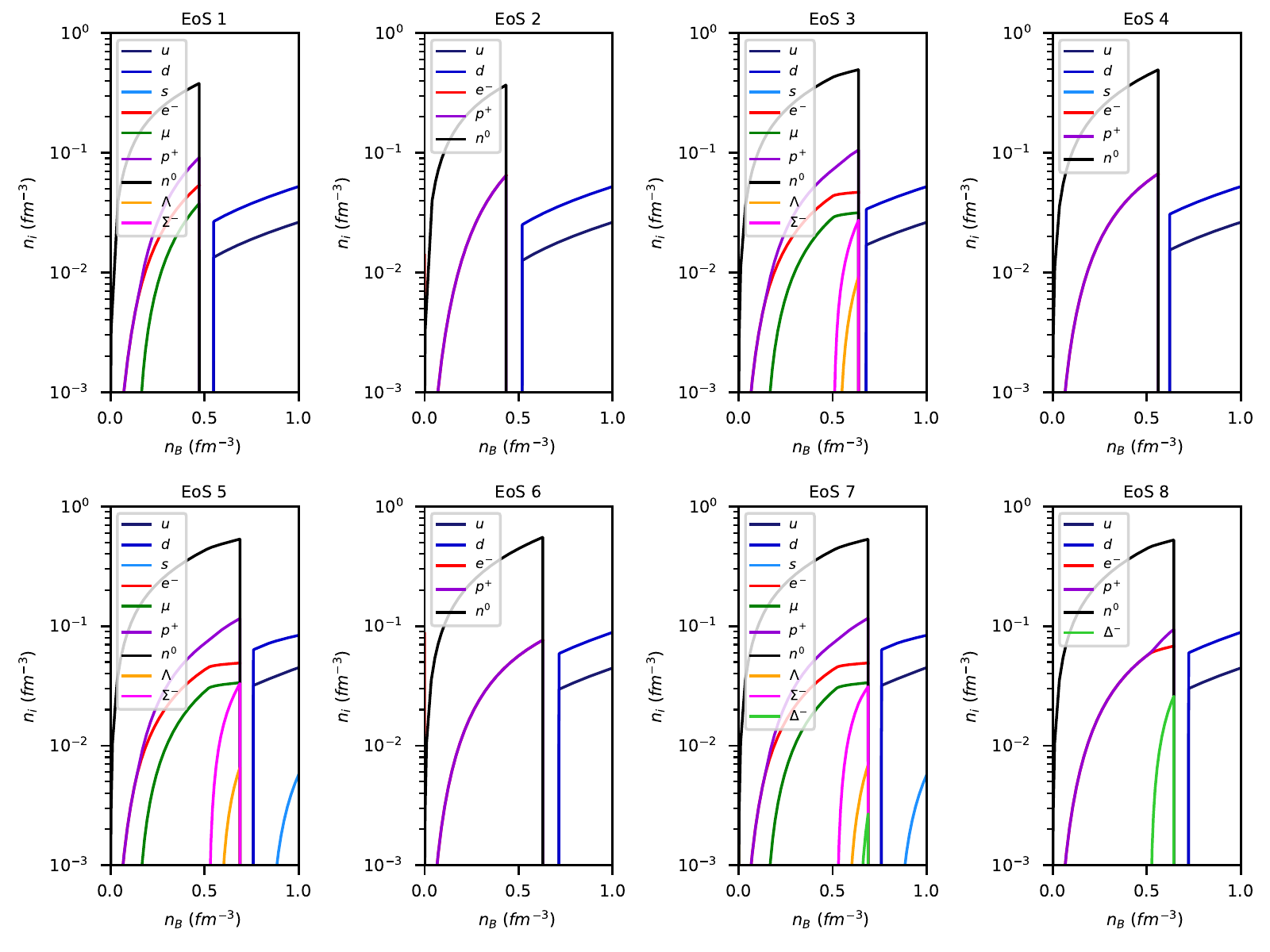}
  \caption{Particle population as a function of density for different hybrid equations of state. For EoS's 2, 4, and 6, the line for the electrons coincides with the one for the protons. Quark densities are divided by $3$.}
  \label{fig:2}
\end{figure*}

Figure \ref{fig:2} shows how the position of the phase transition (density at which it happens), size (extent of jump in density), and quark amount depends on the particle population and interactions included. The combination of the $\omega\rho$ and $\omega^4$ interactions pushes the phase transition to higher densities (EoS 5 and 6 in comparison to EoS 3 and 4, in comparison to EoS 1 and 2). This has to do with how vector and isovector interactions stiffen/soften each phase differently. The absence of hyperons and muons (and strange quarks) modifies the density at which the phase transition happens and makes it stronger (larger extent of density), which can be seen when comparing even (2, 4, 6, and 8) with odd numbered (1, 3, 5, and 7) EoS's. The strange quarks, when included (in the EoS's with hyperons), never appear in larger amounts in the interval of density studied, given their larger bare mass.

In order to include the effects of nuclei in the EoS, we add a zero-temperature beta-equilibrated unified crust by Gulminelli and Raduta (see Ref.~\cite{Gulminelli:2015csa} and references therein), also available in CompOSE, to our EoS tables. In the case of standard vector interactions (EoS’s 1 and 2), we consider in the crust the effective interaction Rs proposed by Friedrich and Reinhard \cite{PhysRevC.33.335} and in the case of including the additional $\omega\rho$ vector interactions (EoS’s 3 to 8) we consider the effective interaction SkM proposed by L. Bennour et al. \cite{PhysRevC.40.2834}, both with cluster energy functionals from Danielewicz and Lee \cite{DANIELEWICZ200936}. In this way, we guarantee that the symmetry energy slope of all EoS's does not jump across the crust-core construction. Note that including a crust (specially the outer part) is essential to obtain the correct macroscopic properties for neutron stars.

\begin{figure*}[t!]
  \includegraphics[width=1.01\linewidth]{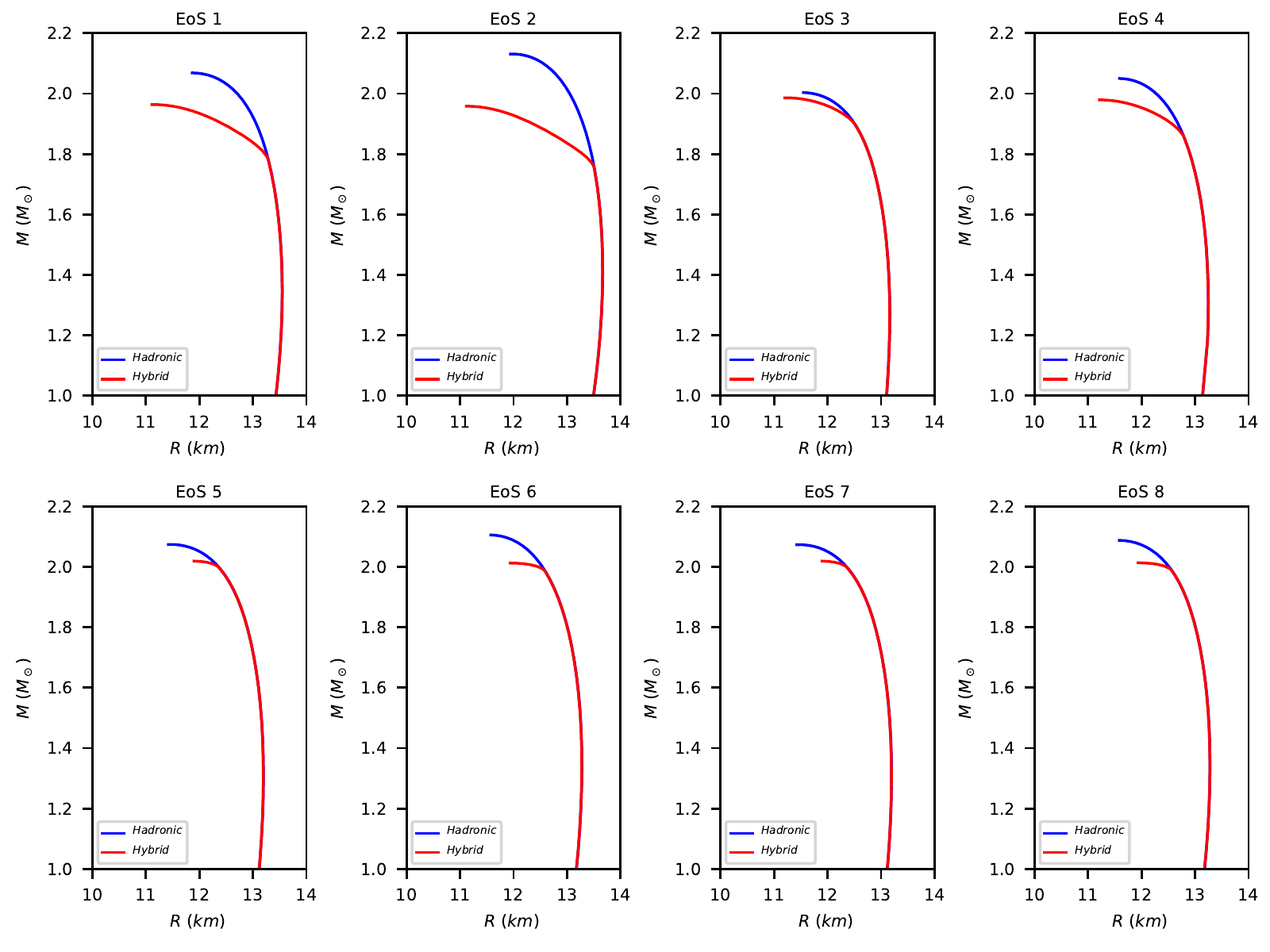}
  \caption{Mass-radius diagram for different hadronic and hybrid equations of state. Only stable stars are shown.}
  \label{fig:3}
\end{figure*}

Solving the Tolman-OppenheimerVolkoff (TOV) equations \cite{Tolman:1939jz,Oppenheimer:1939ne} with our complete EoS's, we obtain families of neutron stars. Stable solutions, to the right side of the maximum mass of each curve \cite{Alford:2017vca}, are shown on Fig.~\ref{fig:3}. The blue curves show how hadronic EoS's with hyperons generate lower maximum-mass stars than EoS's without hyperons (odd in comparison to even numbered hadronic EoS's). This can be seen in Table 1. EoS's with hyperons also generate stars that are slightly smaller, present lower dimensionless tidal deformability \cite{Favata:2013rwa}, and reach larger central densities. The $\Delta$ baryons do not further reduce the maximum-mass of stars generated by EoS 7 (when compared to EoS 5) because they replace hyperons as new degrees of freedom. In the case of EoS 6 with nucleons only, adding $\Delta$'s (EoS 8) decreases (slightly) the maximum mass of stars, but still does not affect other stellar properties (also shown in Table 1). Note that, independently of the composition, the interaction $\omega\rho$ has the effect of decreasing stellar radii and tidal deformabilities (EoS 3 and 4 when compared to EoS 1 and 2) and the interaction $\omega^4$ has the effect of increasing stellar masses (EoS 5 and 6 when compared to EoS 3 and 4).

The red curves in Fig.~\ref{fig:3} show hybrid EoS's. The place where they detach from the (blue) hadronic curves marks the stars in which the central density  surpass the deconfinement density (listed in the Formalism section when the EoS's are listed). In all cases, this takes place in stars that are less massive than the maximum-mass star of the family. EoS's 1 and 2 present longer hybrid branches, but in turn reproduce less massive hybrid stars, just below $2$ M$_\odot$. EoS's 3 and 4 also reproduce hybrid stars just below $2$ M$_\odot$, but now due to the introduction of the $\omega\rho$ interaction, which on the other hand turns all stars of the family smaller.  Finally, EoS's 5, 6, 7, to 8 produce hybrid stars with masses above $2$ M$_\odot$. Note that, while the range of maximum masses for hybrid stars varies only by $0.06$ M$_\odot$, the stellar masses corresponding to the entire hybrid branch change considerably when comparing different couplings and compositions in the top and bottom rows of Fig.~\ref{fig:3}.

\begin{table}[t!]
\caption{Relevant neutron-star properties for different hadronic and hybrid equations of state: maximum mass, corresponding radius and central density, radius and dimensionless tidal deformability of $1.4$ M$_{\odot}$ star.}
\label{tab:1}       
\begin{tabular}{|l|l|l|l|l|l|}
\hline
EoS&M$_{\rm{max}}$&R of M$_{\rm{max}}$&${\rm{n}_{\rm{B}}}_{\rm{c}}$ of M$_{\rm{max}}$&R of $1.4$ M$_{\odot}$&$\tilde{\Lambda}$ of $1.4$ M$_{\odot}$\\
 &(M$_{\odot}$)&(km)&(fm$^{-3}$)&(km)& \\
\hline
1 hadronic & 2.07 & 11.87 & 0.916 & 13.55 & 889  \\
\hline
1 hybrid & 1.96 & 11.11 & 1.079 & 13.55 & 889 \\
\hline
2 hadronic & 2.13 & 11.95 & 0.751 & 13.67 & 904 \\
\hline
2 hybrid & 1.96 & 11.11 & 1.079 & 13.67 & 904 \\
\hline
3 hadronic & 2.00 & 11.55 & 0.964 & 13.15 & 702 \\
\hline
3 hybrid & 1.99 & 11.20 & 1.040 & 13.15 & 702 \\
\hline
4 hadronic & 2.05 & 11.59 & 0.956 & 13.24 & 739   \\
\hline
4 hybrid & 1.98 & 11.21 & 1.040 & 13.24 & 739 \\
\hline
5 hadronic & 2.07 & 11.42 & 0.988 & 13.18 & 723\\
\hline
5 hybrid & 2.02 & 11.89 & 0.892 & 13.18 & 723 \\
\hline
6 hadronic & 2.11 & 11.58 & 0.946 & 13.27 & 754  \\
\hline
6 hybrid & 2.01 & 11.94 & 0.892 & 13.27 & 754 \\
\hline
7 hadronic & 2.07 & 11.42 & 0.988 & 13.18 & 723 \\
\hline
7 hybrid & 2.02 & 11.90 & 0.892 & 13.18 & 723\\
\hline
8 hadronic & 2.09 & 11.58 & 0.950 & 13.27 & 754 \\
\hline
8 hybrid & 2.01 & 11.94 & 0.892 & 13.27 & 754\\
\hline
\end{tabular}
\end{table}

\section{Discussion and Conclusions}

The CMF model offers an ideal mechanism to produce equations of state (EoS's) for astrophysical purposes: it has been fitted to be in agreement with low- and high-energy physics data; can be applied at zero, as well as intermediate and large temperatures; possess the degrees of freedom expected to appear in different astrophysical scenarios (leptons, baryons, quarks) within one description; reproduces QCD features, such as chiral symmetry restoration and deconfinement to quark matter; it is relativistic, therefore, presenting causal behavior. Note that very strong repulsive terms could violate causality, but this is not the case in our work (see discussion in footnote of Ref.~\cite{Dexheimer:2020rlp}).

The $16$ EoS's we discuss in this work vary i) according to composition: with hyperons (plus s-quarks) and muons vs. with only nucleons (plus u,d-quarks) and electrons, ii) hadronic vs. hybrid, and iii) with vs. without different vector interactions. The mixed vector-isoscalar vector-isovector $\omega\rho$ interaction softens the matter EoS at low densities, while the higher-order vector-isoscalar interaction stiffens the matter EoS at large densities.

The 8 new hybrid EoS's discussed in the previous section are available as one-dimensional EoS tables in the CompOSE repository \cite{compose}, together with the 8 corresponding hadronic EoS's, with the (baryon number) density as the independent variable. Our numerical code runs as a function of baryon chemical potential, a necessary procedure when more than one phase exists. This eliminates the possibility of finding solutions that are not truly stable (without  globally minimized free energy). The data produced by our code is then linearly interpolated to appear in CompOSE as a function of density and at regular intervals of density. Tables with thermodynamical properties, composition of matter, stellar information, and microscopic information are provided, with and without crust. See the summarized CompOSE manual for details on the available files \cite{providers}. 

From our analysis, whenever more massive baryons, hyperons and $\Delta$'s, are allowed to appear, they do so before the phase transition to quark matter, which is relatively weak and, therefore, allows for hybrid stars to be stable. All hadronic and hybrid EoS's discussed produce stars with masses around or above $2$ M$_{\odot}$, consistent with observations of massive stars. In our work, strange quarks only appear (when included) in very small amounts in massive hybrid stars. As a consequence, observable effects of strangeness, such as the one performed in Ref.~\cite{Tan:2021ahl}, based on significant changes in the speed of sound of matter associated with the appearance of new degrees of freedom, would have to be based on the hadronic phase only, where there is enough strangeness for this to manifest. Observables for the phase transition itself would be hard to find, given how weak it is. While most of our hybrid stars fall into the category of masquerade stars \cite{Alford:2004pf,Cierniak:2021knt}, some of them present a significant kink. The significant change in slope in the mass-radius plane associated with these kind of kinks have recently been associated with observable effects in the binary Love relations \cite{Tan:2021nat}. Work on neutron-star merger simulations with the EoS's presented here is already underway and will appear in a future publication, as well as finite temperature versions of the EoS's discussed in this work.

\bmhead{
We acknowledge support from the National Science Foundation under grants PHY1748621, MUSES OAC-2103680, and NP3M PHY-2116686, in addition to PHAROS (COST Action CA16214).}

\bibliography{mainbib}
\end{document}